# Scanning Tunneling Microscope Nanolithography on SrRuO$_3$ Thin Film Surfaces

Yun LIU[1)], Jia ZHANG[2)]

1) Department of Physics, Norwegian University of Science and Technology, Trondheim 7491, Norway
2) School of Mechanical Engineering, University of South China, Hengyang, 421001, China

Nanoscale lithography on SrRuO$_3$ (SRO) thin film surfaces has been performed by scanning tunneling microscopy under ambient conditions. The depth of etched lines increases with increasing bias voltage but it does not change significantly by increasing the tunneling current. The dependence of line width on bias voltage from experimental data is in agreement with theoretical calculation based on field-induced evaporation. Moreover, a three-square nanostructure was successfully created, showing the capability of fabricating nanodevices in SRO thin films.

## 1. Introduction

The feasibility of manipulating the material surface in nanometer length scale on demand by scanning probes was first demonstrated by Becker et al.[1] in 1987. Since then much effort has been devoted to this rapidly developing area, later named scanning probe lithography (SPL)[2-8]. Currently, SPL-based technology has attracted much attention in creating nanoscale structures on a variety of materials while conventional lithography methods are approaching fundamental size limits[9]. Particularly in modern miniaturization of microelectronic, SPL has offered promising possibilities in the fabrication of nanoscale devices and development of engineered templates for growth of epitaxial thin films.

Many materials surfaces such as silicon[4,6], graphite[10,11], and YBa$_2$Cu$_3$O$_7$[12-14], have been modified by SPL. The creation of nanostructures was found to be strongly influenced by bias voltage, tunneling current, scan speed, and ambient conditions. The lithography mechanism varies depending on experimental conditions and is still debated. Among others, field-induced evaporation[15] is widely accepted as a critical or in some cases dominant scheme in the context of lithography experimentation[16-18]. That is, when the tip-sample proximity decreases to several angstroms, the applied electrical field becomes strong enough to break up the atomic bonding, and atoms are ionized and evaporate away from the sample or tip surfaces. The field-induced evaporation is a thermally activated process and the rate of evaporation is given by $\kappa = v\exp(-Q/kT)$, where $Q$ is the activation energy and $v$ is the frequency factor (which is about $10^{13}$ s$^{-1}$). Field evaporation theory has successfully explained many phenomena discovered in SPL experiments such as threshold bias voltages and reversibility of material transfer[4].

In the present work, we have performed scanning tunneling microscope (STM) lithography on SrRuO3 (SRO) thin films surfaces under ambient conditions using self-fabricated iridium (Ir) tips. SRO is a conductive perovskite material which has an orthorhombic structure with the space group Pbnm and lattice parameters $a$=5.5670Å, $b$=5.5304Å and $c$=7.8446Å [19]. Epitaxial functional oxide thin film based on the SRO template could be crucial for the application of nanoelectronic devices.

## 2. Exprimental method

SRO (110) thin film was epitaxially grown on (001)-oriented SrTiO3 (STO) substrate by off-axis radio frequency magnetron sputtering. STM measurement revealed its step-and-terrace topography with step heights of one to two unitcells. The rms roughness of the terraces surfaces was found to be ~1 Å. X-ray diffraction analysis indicated excellent crystalline quality, the full width at half maximum (FWHM) of rocking curve around the (110) peak was determined to be less than 1°. STM tips were prepared from Φ0.25 mm Ir wire by electrochemical method. Ir tips with macroscopic radius of curvature (ROC) ~50 nm were routinely produced; the detailed procedure will be described in a separate publication.

The nanoscale lithographic experiments were conducted on a commercial air STM system at room temperature. The images of modification patterns were obtained right after the lithographic process by the same tips. Both imaging and lithography operations were under constant current mode and with feedback loop switching on. Scanning parameters for normal imaging were positive tip bias voltage of 500 mV and tunneling current of 500 pA; for lithography, the etching parameters were positive bias voltage ranging from 1.8 V to 2.5 V, 60 pA setpoint tunneling current, 500 nm/s scan speed, and 100 scan repetitions. During the lithographic process, the tip movement was defined by a Nanoscript$^{TM}$ programme. Fig.1 demonstrates an eight-line pattern fabricated by scanning tips over selected area. Iridium tips scanned back and forth to create a 100 nm long single line. The upper left line was first etched and the lower right last. The sequence is indicated by the numbers in Fig.1. It should be noted that in this experiment only bias voltage that was applied positive to iridium tips offered successful line-etching on SRO thin film surfaces.

## 3. Results and Discussion

As can be seen from Fig.1, there is an evolution process from line 1 to line 4 in the upper row. After four lines, the etching is more stable and reproducible. Fig.2 shows the dependence of line depth on etching sequence. The two end parts of each line were removed in the depth determination because there is a time lag when the tip is changing direction at the ends of each line, leading to deeper etching. Additionally, we define a successful etching if etched line is continuous for at least 70 nm with a minimum average depth of half a unit cell. Therefore, only successful lines from the bottom row were analyzed in this work.

The analyzed line depth as a function of bias voltage and tunneling current is plotted in Fig.3. The line depth increased from 0.6 nm to 7.7 nm as bias voltage increasing from 1.8 V to 2.8 V, while the increase of the tunneling current does not change the depth significantly. This observation therefore suggests that etching mechanisms such as electronmigration[2] and local heating[17] could be ruled out because both are strongly current dependent. The slow increase of depth with tunneling current is consistent with field-induced emission in which the electric field $F$ is related to tunneling current $I$ by $F \propto -1/\log I$[6]. In other words, the field induced evaporation is probably a dominant process in the STM lithography with a combination of the Ir tip and SRO surface. However, as bias voltage is larger than 2.8 V, tip behavior was easily altered, resulting in unstable etching into shallow double lines due to the formation of double tips.

In order to verify the field-induced mechanism, Fig.4 gives the dependence of line width on bias voltage obtained from both experimental data and theoretical calculation. The tip apex shape can be modeled in a prolate-spheroidal coordinate system[20,21].

The three orthogonal coordinates ($\xi, \eta, \phi$) are defined by

$$\begin{cases} x = a\sqrt{\xi^2-1}\sqrt{1-\eta^2}\cos\phi, \phi \in [0, 2\pi] \\ y = a\sqrt{\xi^2-1}\sqrt{1-\eta^2}\sin\phi, \xi \in [1, \infty] \\ z = a\xi\eta, \eta \in [\eta_1, \eta_2] \end{cases} \quad (1)$$

Where $a$ is one-half distance between the hyperboli foci; $\phi$ is the azimuthal angle around the central axis on which the foci are located; $\xi$ and $\eta$ describe prolate spheroids and hyperboloid of revolution respectively; $\eta_1$ corresponds to the tip surface and $\eta_2$ corresponds to the SRO surface. Here, we focus on the region between two hyperboloid surfaces which is analogous to the physical situation in STM. For the tip-sample system, we also have the following formulas[21]

$$\begin{cases} a\cos\theta = d \\ \tan\theta = \sqrt{r/d} \end{cases} \quad (2)$$

In Eq. (2), $r$ is tip ROC, $d$ is tip-sample spacing and $\theta$ is a parameter determined by $r$ and $d$. On the tip surface we have $\eta = \cos\theta$. The electric field $F$ between STM tip and sample surface can be expressed as [20].

$$F = \frac{2V\tan\theta}{r\ln\frac{1+\cos\theta}{1-\cos\theta}\sqrt{\xi^2 - \cos^2\theta}} \quad (3)$$

We define the minimum electric field for evaporating SRO surface atoms to be $F_c$. From Eq.(1) and (3), the field emission radius $R$ on the SRO surface is given by

$$R = \frac{d\sqrt{\cos\theta}}{\cos\theta} \sqrt{\frac{V^2 \tan^2\theta}{r^2 F_c^2 \ln^2 \frac{1+\cos\theta}{1-\cos\theta}} - \sin^2\theta} \qquad (4)$$

After substituting Eq.(2) in Eq.(4), $R$ is ultimately determined by the bias voltage $V$. The tip-sample spacing was evaluated to be varying between 0.5 to 1 nm by taking the work function of bulk SRO 5.2 eV to be the tunneling barrier[22]. The threshold bias voltage for etching SRO thin film surfaces using Pt/Ir tips was determined to be 1.6 V by extrapolating the guide line to the y axis in the plot of bias voltage against scan speed[23]. Considering the tip-sample spacing mentioned above, the critical electric field of SRO surface atoms could be estimated to range from 1.6 to 3.2 V/nm. By choosing tip radius 50 nm, tip-sample distance 0.5 nm, and minimum evaporation field for SRO 1.7 V/nm, the calculated curve correlating line width $2R$ and bias voltage $V$ fits very well with the experimental data.

Fig.5 illustrates the spatial distribution of the electric field on the SRO surface. This plot is based on Eq.(3) and the tip is modeled as a 50 nm radius sphere 0.5 nm above a semi-infinite metal surface. It is found that the electric field is concentrated in the region under the tip apex, and only the SRO surface area determined by emission radius is modified during the STM lithographic process.

We have also created complex structures to examine the feasibility of fabricating nanodevices on SRO thin film surfaces. Fig.6 (a) shows a three-square structure created by using bias voltage from 2.2 V to 2.4 V, tunneling current 60 pA, scan speed 500 nm/s and scan repetition 100 per line. The section profile in Fig.6 (b) corresponds to the white marker line in Fig.6 (a). The reproducibility of etching complicated structures is heavily dependent on the tip quality. In this work, on average one out of five Ir tips is able to sustain through etching of one three-square structure.

## 5. Conclusion

In summary, STM nanolithography has been performed on SRO thin film surfaces. The line depth increases with increasing bias voltage but does not change significantly with increasing tunneling current. The dependence of width on bias voltage is in good agreement with theory based on field-induced evaporation. Moreover, a three-square structure has been created to show the feasibility of fabricating electronic nanodevice on SRO thin film surfaces.

**Figure captions**

**Fig.1** STM image (400nm×400nm) of eight etched lines from the line etching experiments. Each line was 100 nm long and etched at bias voltage 2.4 V, tunneling current 60pA, scan speed 500 nm/s and 100 scan repetitions per line.

**Fig.2** Depth vs. etching sequence. Etching parameters are bias voltage 2.6 V, tunneling current 60 pA, scan speed 500 nm/s and 100 scan repetitions per line.

**Fig.3** Dependence of line depth on both bias voltage and tunneling current using scan speed 500 nm/s and scan repetition 100. The tunneling current for depth-bias voltage measurement is 60 pA.

**Fig.4** The line width as a function of the bias voltage from both experimental data and field evaporation theory.

**Fig.5** Spatial distribution of electric field on SRO thin film surface.

**Fig.6** (a) Three-square structure created by STM lithography. The bias voltage for the large square is 2.4 V, for the medium one, 2.2 V, and for the small one, 2.0 V. Other parameters are tunneling current 60 pA, scan speed 500 nm/s, and scan repetition 100. (b) The section profile along white marker line in (a).

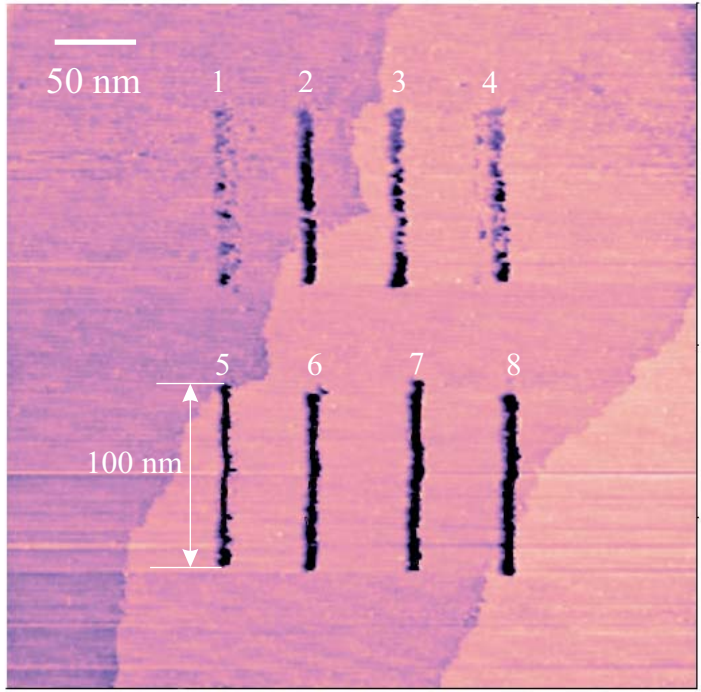

Fig. 1

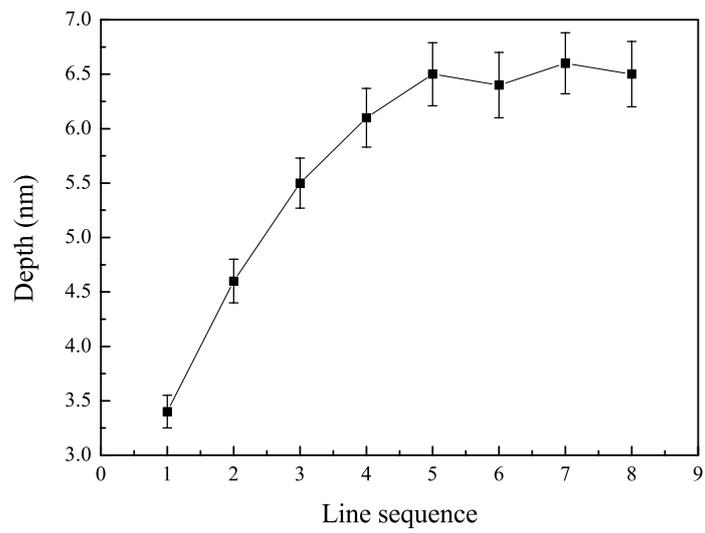

Fig. 2

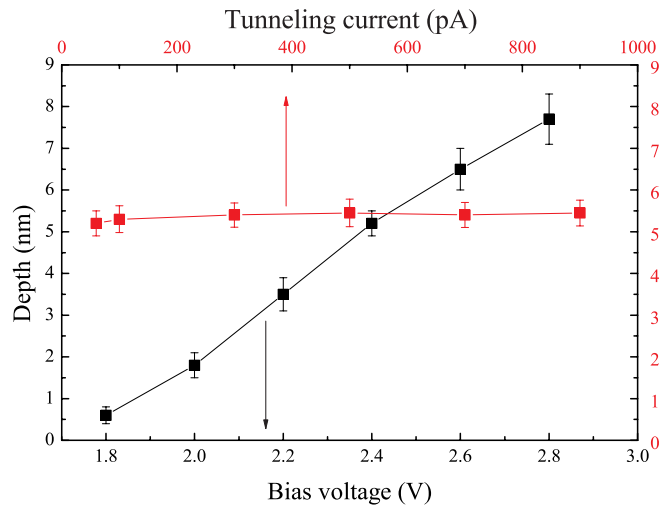

Fig. 3

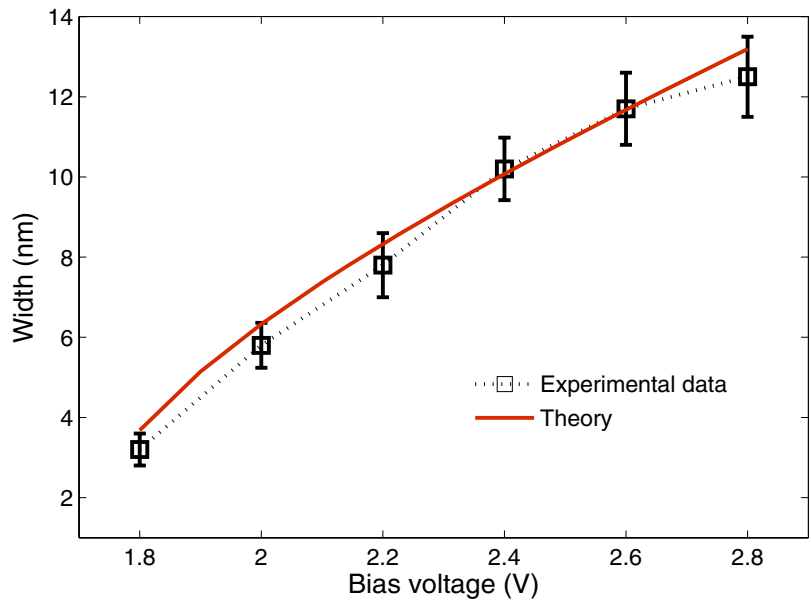

Fig. 4

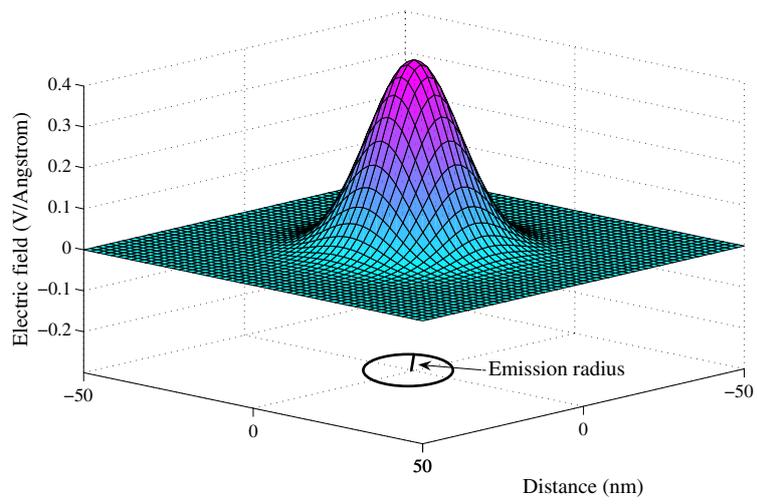

Fig. 5

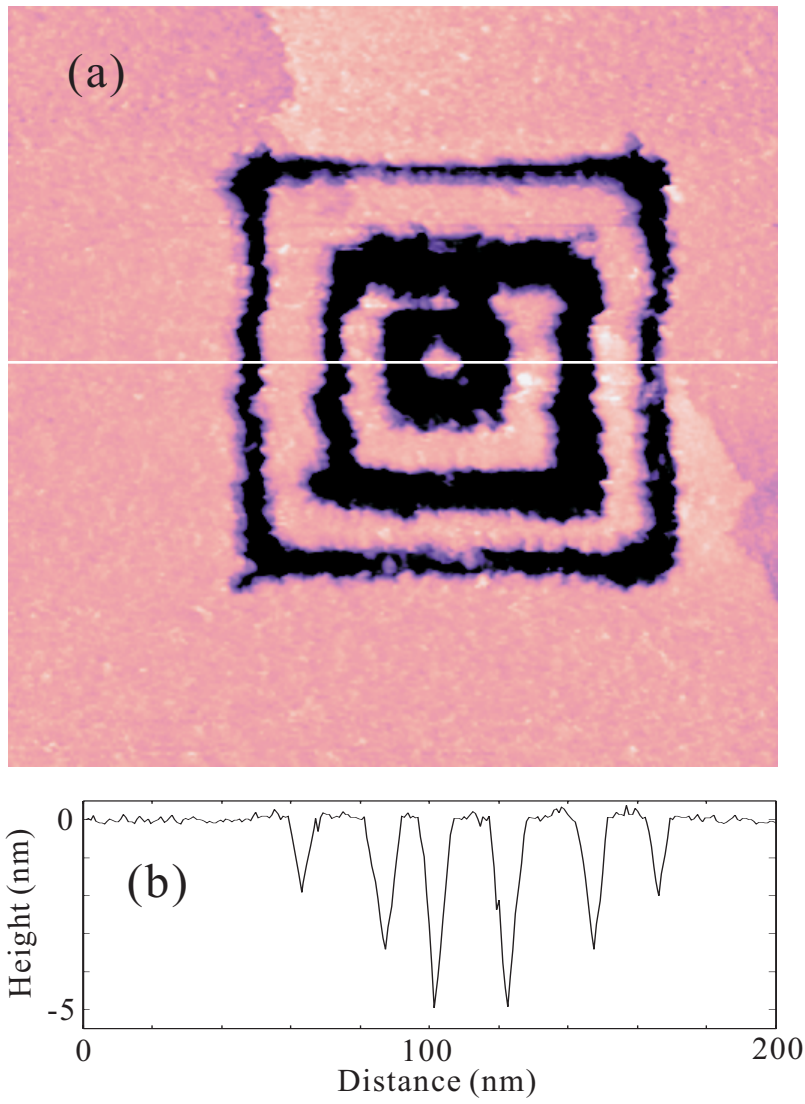

Fig. 6